\documentclass[aps,prl,amssymb,twocolumn,superscriptaddress]{revtex4-1}
\usepackage{graphicx}
\usepackage{amsmath}
\usepackage{dcolumn}
\usepackage{bm}
\usepackage{color}
\usepackage{xcolor}
\UseRawInputEncoding

\newcommand{\ocl}{\text{OC}_1}
\newcommand{\ocr}{\text{OC}_2}
\newcommand{\ocj}{\text{OC}_j}
\newcommand{\ml}{\text{WGM}_1}
\newcommand{\mr}{\text{WGM}_2}
\newcommand{\mj}{\text{WGM}_j}
\newcommand{\quoting}[1]{``#1''}
\newcommand{\eq}[1]{Eq.~(\ref{#1})}

\begin{document}

\title{
Impact of non-Hermitian Mode interaction on Inter-cavity Light transfer
}

\author{Hyeon-Hye Yu}
\author{Sunjae Gwak}
\author{Jinhyeok Ryu}
\author{Ji-Hwan Kim}
\affiliation{Department of Emerging Materials Science, DGIST, Daegu 42988, Republic of Korea}
\author{Jung-Wan Ryu}
\affiliation{Center for Theoretical Physics of Complex Systems, Institute for Basic Science (IBS), Daejeon 34126, Republic of Korea}
\author{Chil-Min Kim}%
 \email{chmkim@dgist.ac.kr}
\author{Chang-Hwan Yi}%
 \email{innissan@dgist.ac.kr}
\affiliation{Department of Emerging Materials Science, DGIST, Daegu 42988, Republic of Korea}


\begin{abstract}
  Understanding inter-site mutual mode interaction in coupled physical systems is essential to comprehend large compound systems as this local interaction determines the successive multiple inter-site energy transfer efficiency. We demonstrate that only the non-Hermitian coupling can correctly account for the light transfer between two coupled optical cavities. We also reveal that the non-Hermitian coupling effect becomes much crucial as the system dimension gets smaller. Our results provide an important insight to deal with general coupled-devices in the quantum regime.
\end{abstract}

\maketitle

    Unit pairwise coupled systems have been an essential subject in almost all research fields of physics, as they serve the most foundational element to constitute the larger-scale complex of physical systems in nature. Up to date, numerous coupled systems in various intrinsic physical states have been implemented, such as hybrid quantum information systems~\cite{p1qi1,p1qi2,p1qi3}, optical communication systems~\cite{p1oc1,p1oc2,p1oc3}, and topological photonics systems~\cite{p1to1,p1to2,p1to3,p1to4}. Here, the important property we have to understand for the successful realization of these systems is the inter-site mutual mode couplings.
    
    As the cutting edge of modern technology reaches even up to realizing optoelectronic circuits, a growing interest has been focused on coupled optical microcavities (COCs). So far, a large number of theoretical and experimental results have demonstrated their outstanding potential as an efficient on-chip operational component.
    Because COCs have a powerful capability in manipulating resonant modes and dispersions, they can be used for devices such as optical delay lines~\cite{delayline1,delayline2,delayline3}, filters~\cite{filter1,filter2,filter3,filter4}, and switches~\cite{swit1,swit2,swit3}.
    Recently, they are considered as promising candidates in more high-tech future applications, e.g., optical memory~\cite{mem1,mem2,mem3,mem4}, highly sensitive sensors~\cite{sen1,sen2,sen3}, and single-mode lasers~\cite{sinmod1,sinmod2,sinmod3}. Add more, it is also emphasized that they can perform the broad-ranged mode couplings from the near field evanescent regime to the ultra-long distance coupling~\cite{broad1,broad2}. Currently, COCs have attracted newly stimulated interests associated with non-Hermitian physics, such as the non-Hermitian degeneracy, so-called exceptional points(EPs)~\cite{Kato,Heiss,Peng}, the parity-time symmetry~\cite{Bender,El,Lan,PT4,PT5,PT6,PT7}, and the photonic molecular states~\cite{PM1,PM2,Boris1,Boris2,PM3,PM4,PM5,PM6,PM7} because of their intrinsic openness property~\cite{Okoowicz,Hentschel}.
    
    Generally, COCs include waveguide structures that have the role of external input-output ports of light. There are two independent coupling processes in these systems: (i) waveguide $\Leftrightarrow$ cavity and (ii) cavity $\Leftrightarrow$ cavity. Here, we have to point out that most literature focused only on the case (i)~\cite{g2c1,g2c2,g2c3,g2c4} so far. Even for the studies considering case (ii), a lossless coupling between cavities are typically assumed~\cite{c2c1,c2c2,c2c3,c2c4}. Strictly speaking, this "artificial" assumption is insufficient for describing real physical incidences. The inter-cavity mode coupling belonging to the case (ii) requires more careful treatment to account for it than the case (i). It is because the mutual feedback of light transfer between the cavities induces more complicated interactions. Furthermore, as the coupling occurs via the free space between the cavities, it should involve external coupling effects, as well~\cite{Wiersig2006}.
    
    \begin{figure}[b]
	\includegraphics[width=0.85\columnwidth]{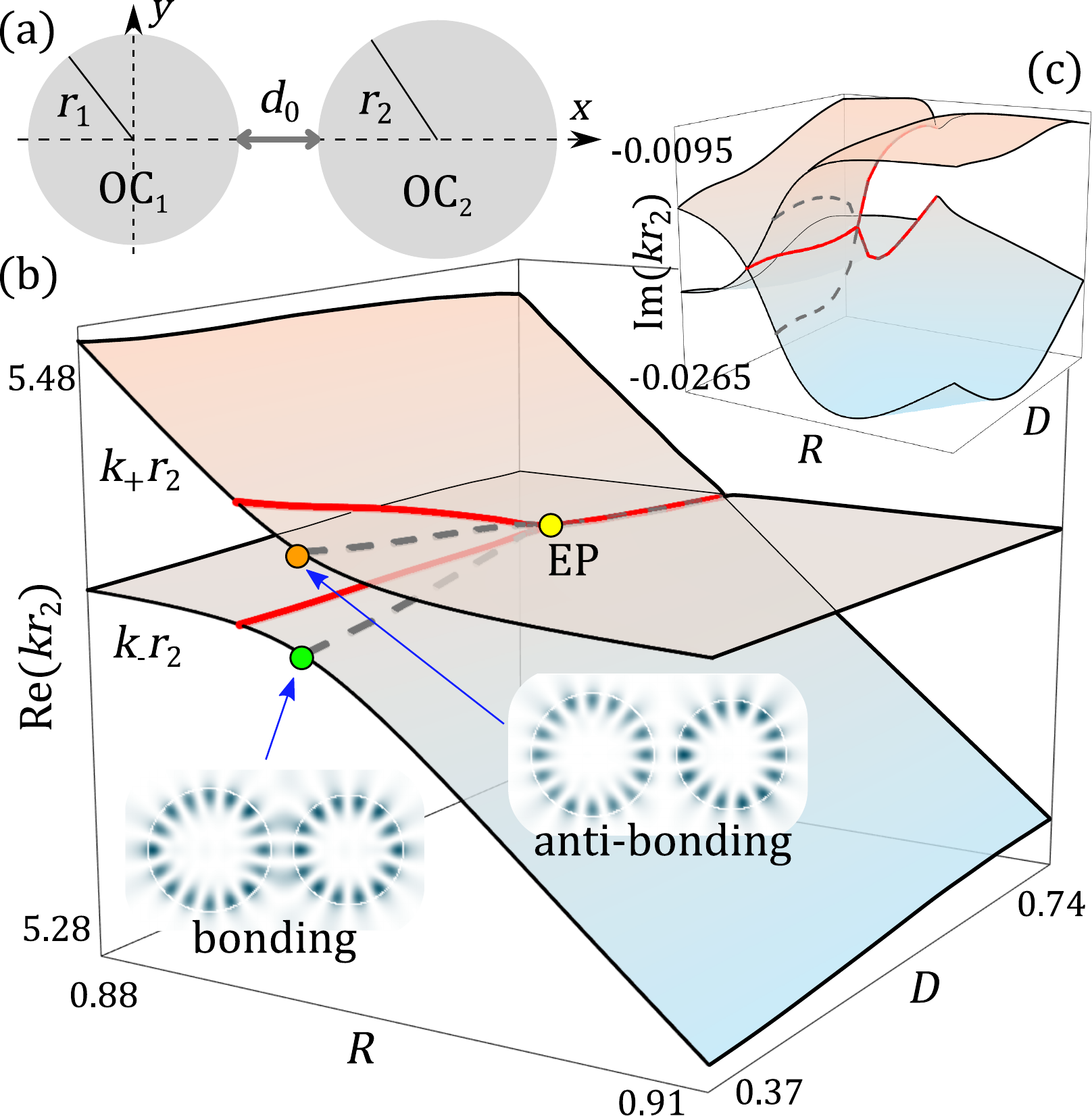}
	\caption{(a) is the system configuration of coupled microcavities, where $r_1$ and $r_2$ are radii of each cavities, and $d_0$ is the inter-cavity distance. (b) and (c) are the Riemann surfaces for the real and imaginary parts of resonant wavenumbers in parameter space $(R,D)$, obtained by BEM. The branch-cut (the interaction center) is marked by red solid (gray dashed) curve. Insets in (b) are the spatial distribution of two coupled modes at the interaction center.}
	\label{sys}
\end{figure}
    In this Letter, we will show that inter-cavity light transfer in the strong-coupling regime~\cite{Heiss} only can be explained correctly by a lossy coupling between modes. As this coupling results in a fully-non-Hermitian Hamiltonian for resonant modes~\cite{Wiersig2006}, we refer to this lossy coupling as \textit{non-Hermitian coupling} and a lossless coupling as \textit{Hermitian coupling}. To explicitly demonstrate the impact of this non-Hermitian coupling in inter-cavity light transfer, we model two interacting whispering gallery modes (WGMs), each confined in different microdisks. Through time-dependent numerical experiments using the finite difference time domain (FDTD) method, real experimental situations of inter-cavity light transfer are simulated. The results are analyzed by exact numerical results of the boundary element method (BEM)~\cite{WiersigBEM}, as well as the temporal coupled-mode theory (TCMT)~\cite{Haus,Fan}. Our results reveal the fact that the non-Hermitian coupling has a critical impact on the inter-cavity light transfer efficiency, particularly when the cavity sizes approach the quantum regime.

    Figure~\ref{sys}(a) illustrates our COCs, where two dielectric microdisks having radii $r_1$ and $r_2$ are positioned at the distance $d_0$. We set the refractive index $n = 2.0$ inside the disks and $n=1.0$ for outside. We focus on transverse magnetic (TM) polarized WGMs. The insets in Fig.~\ref{sys} show two coupled modes that consist of two basis modes, $\ml$ and $\mr$ confined in each isolated $\ocl$ and $\ocr$. Here, $\ml$ and $\mr$ are defined by $(l,m) =(1,7)$ and $(1,8)$, where $l$ and $m$ represent radial and angular mode numbers, respectively. Throughout this Letter, $r_2$ remains constant, while two parameters $R\equiv r_1/r_2$ and $D\equiv d_0/r_2$ are used for constructing Riemann-surface: solution sets, $\{k\}\in\mathbb{C}$ (wavenumbers), of the Helmholtz equation, $-\nabla^2\psi(r)=n^2(r)k^2 \psi(r)$, in parameter space. Figures~\ref{sys}(b) and (c) examine Riemann-surface of coupled modes around $\text{Re}(kr_2)=5.5$ in $R\in[0.88,0.91]$ and $D \in [0.37,0.6]$, studied in~\cite{JWRyu}. 

\begin{figure}[t]
  \includegraphics[width=0.95\columnwidth]{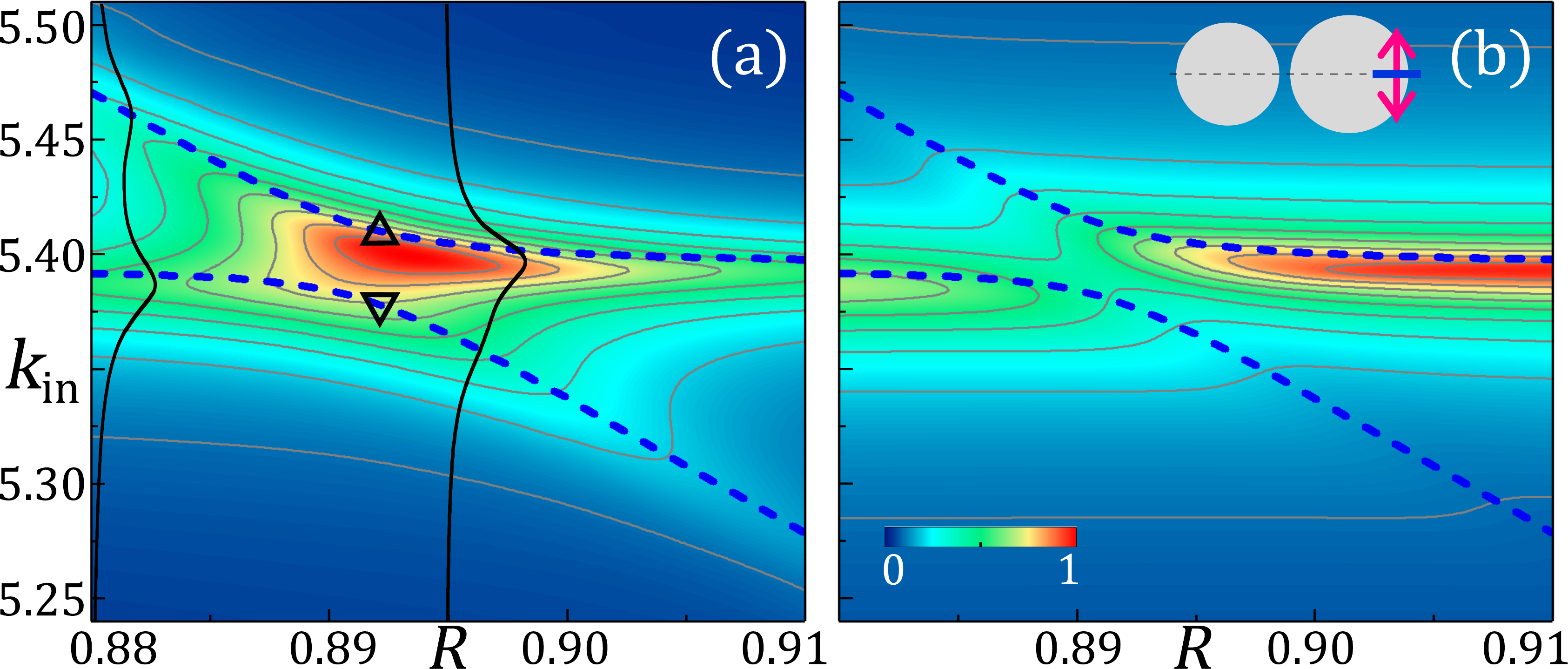}
    \caption{
     (a) is the FDTD results of EDA spectra of $|a_1|$ and (b) of $|a_2|$ at $D=0.37$. Dashed curves are Re($k_\pm r_2$) obtained by BEM. In (a), solid curves at $R=0.88$ and $0.892$ are selected examples of EDA, and the upper/lower triangles mark the bonding/anti-bonding modes. The inset in (b) shows the radiating pumping source (arrow) used in FDTD simulation.
        }
  \label{stddensity}
\end{figure}

    We begin with FDTD calculations for a energy density of steady-state (i.e., $t\to\infty$) electromagnetic fields in each of $\ocl$ and $\ocr$. In this numerical experiment, COC is excited by a TM polarized input source, $e^{-i \omega t}$, located only in $\ocr$, where $i=\sqrt{-1}$, $\omega=c k_\text{in} \in\mathbb{R}$, and $c$ is the speed of light, respectively. The input source radiates bi-directionally along the cavity boundary (see the arrows in the inset of Fig.~\ref{stddensity}) in order to excite the even-parity modes (see insets in Fig.~\ref{sys}) with respect to the horizontal axis. Figure~\ref{stddensity} shows $R$-dependent energy density amplitude (EDA) defined~\cite{Haus} by
\begin{align}
  |a_j(k_\text{in}r_2)| \equiv \bigg[\frac{1}{2A_j}\int_{\ocj} (n^2(\mathbf{r}) |E_j(\mathbf{r})|^2 + |B_j(\mathbf{r})|^2) ~d\mathbf{r}~\bigg]^{\frac{1}{2}}\nonumber \ ,
  \label{amp}
\end{align}
    for a fixed $D=0.37$, where $A_j$ is a disk area of $\ocj$ and $E_j$ ($B_j$) the electric (magnetic) field inside $\ocj$. The integral domain is restricted to the inside the disk $\ocj$.
    
    The energy density amplitude, $|a_1|$, in $\ocl$ can be interpreted as light transfer efficiency from $\ocr$, as only the latter embeds the light source in it. In Fig.~\ref{stddensity}, it is found that the most strong light transfer takes place around a \quoting{interaction center} [$\equiv$ a point where a gap between Re($k_+r_2$) and Re($k_-r_2$) becomes the smallest, see Fig.~\ref{sys}]. At this point, the two interacting modes form the well-known doublet of the bonding and the anti-bonding photonic-molecular states~\cite{PM1}(see, Fig.~\ref{sys}), and they respectively correspond to the lower $(\triangledown)$ and upper $(\vartriangle)$ triangles in Fig.~\ref{stddensity}(a). In the figure, we can identify that the light transfer associated with the anti-bonding mode is much stronger than the bonding mode.

    To clarify the origin of this phenomena, the EDA spectra are analyzed by two-mode TCMT model of weakly coupled COC. In our model, a continuous-wave light source $\sqrt{g_s}S_0e^{-i\omega t}$ oscillates with a frequency $\omega\in\mathbb{R}$ at the cavity $\ocr$:
\begin{equation}
    \begin{aligned}
        \frac{da_1}{dt}&=-i\omega_1 a_1-g_1 a_1 -i\gamma_{12}a_2\\
        \frac{da_2}{dt}&=-i\omega_2 a_2-g_2 a_2 -i\gamma_{21}a_1+\sqrt{g_s}S_0e^{-i\omega t} \ ,
    \end{aligned}
\label{cmeS}
\end{equation}
    where $\omega_j\equiv c \text{Re}(k_j)$ is the resonant frequency and $g_j\equiv c |\text{Im}(k_j)|$ the decay rate of $\mj$, and $\{a_j,\ S_0,\ \gamma_{ij}\}\in\mathbb{C}$ are the mode amplitude, the source amplitude, and the coupling coefficient, respectively. There are two assumptions in our model: the perfect source-to-cavity coupling, i.e., $g_s=1$, and the frequency-independent coupling coefficients which is valid when $|\omega_j|\gg |g_j|$~\cite{Haus}. Given the time-harmonic ansatz of solutions, $a_{j}=a_{j}^0e^{-i\omega t}$, we can obtain the steady-state amplitude $a_{j}^0\in\mathbb{C}$, as follows:
\begin{equation}
    \begin{aligned}
        a_1^0(\omega)&=\frac{-i\gamma_{12}\sqrt{g_s}S_0}{\gamma_{12}\gamma_{21}-(\omega-\omega_1+ig_1)(\omega-\omega_2+ig_2)}\\
        a_2^0(\omega)&=\frac{-i\sqrt{g_s}S_0(\omega-\omega_1+ig_1)}{\gamma_{12}\gamma_{21}-(\omega-\omega_1+ig_1)(\omega-\omega_2+ig_2)}\ .
    \end{aligned}
\label{stdampeq}
\end{equation}
    Essentially, these $a_1^0$ and $a_2^0$ correspond to the FDTD results in Fig.~\ref{stddensity}(a) and (b), respectively. Here, we emphasize that the coupling coefficients $\gamma_{ij}$ in \eq{stdampeq} are crucial for reproducing FDTD experiments. Most importantly, it turns out that the typical assumption of a lossless coupling, $\gamma_{12}=\gamma_{21}^*\in \rm \mathbb{C}$ or $\gamma_{12} = \gamma_{21}\in \mathbb{R}$~\cite{Haus} is valid only in the classical limit ($\text{Re}(kr_2) \gg 1$) and may fail as we approach the quantum regime ($\text{Re}(kr_2) \sim 1$).
\begin{figure}[t]
	\includegraphics[width=0.85\columnwidth]{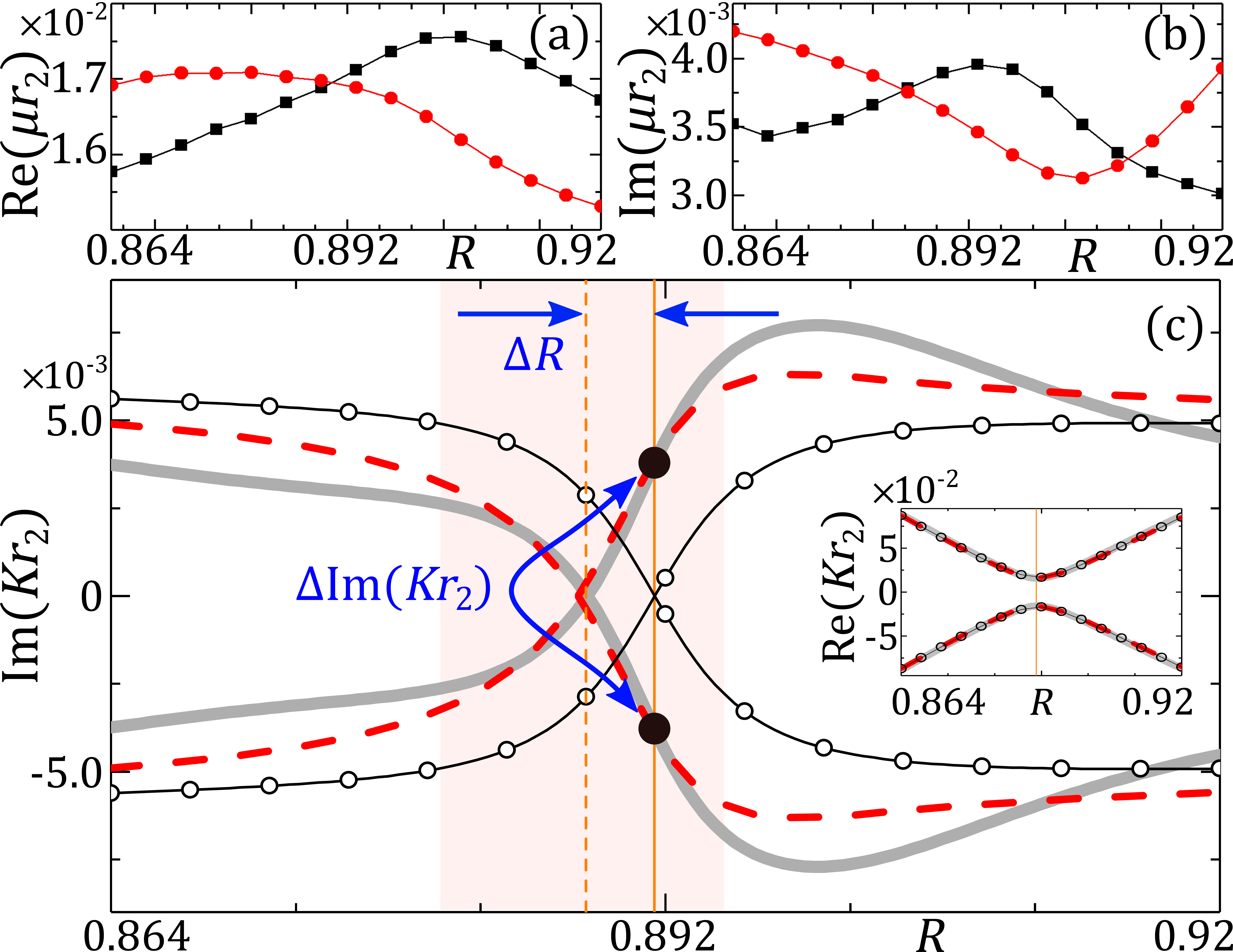}
	\caption{
    Couplings of $\ml$ and $\mr$. Real (a) and imaginary (b) part of $\mu_{12}r_2$ (red-circle) and $\mu_{21}r_2$ (black-square) obtained by \eq{solmu}, as a function of $R$ at $D=0.37$. In (c), Im($K_\pm r_2$) obtained by BEM (gray-solid) are compared to those by \eq{eigen} with the Hermitian (open-circle) and non-Hermitian (red-dashed) couplings. Inset in (c) shows Re($K_\pm r_2$). Here, $K_\pm r_2$ are the re-expressed relative eigenvalues for mean eigenvalues. The vertical solid lines in (c) and the inset mark the interaction center, and dashed line in (c) mark the branch-cut.
     }
	\label{m7m8mu}
\end{figure}

    For the explicit demonstration of this finding, we set the effective Hamiltonian which has complex wavenumber eigenvalues, $k_\pm\in \mathbb{C}$, of the coupled modes, as follows:
\begin{align}
    \mathbf{H}_\text{eff}
    \mathbf{c}^{\pm}
    =
    \begin{pmatrix}
        k_1 & \mu_{12} \\
        \mu_{21} & k_2
    \end{pmatrix}
    \begin{pmatrix}
        c_1^{\pm} \\ c_2^{\pm}
    \end{pmatrix}
    =k_\pm
    \begin{pmatrix}
        c_1^{\pm} \\ c_2^{\pm}
    \end{pmatrix} \ ,
\label{eigen}
\end{align}
    where $k_j\equiv(\omega_j-ig_j)/c$ are complex wavenumbers of $\mj$ with amplitudes $E_j(\mathbf{r})$, and $\mu_{ij}\equiv\gamma_{ij}/c$. The eigenvectors $(c_1^\pm,c_2^\pm)^\text{T}$ are coefficients of the two basis modes and construct new eigenstates of coupled modes; $E_\pm(\mathbf{r})=[c_1^\pm E_1(\mathbf{r})+c_2^\pm E_2(\mathbf{r})]$. For the known $k_\pm$, we can fix $\mu_{ij}$ after obtaining those coefficients numerically. Exploiting the general bi-orthogonality of modes in non-conservative systems, $\int n_i^2(\mathbf{r})E_i(\mathbf{r})\cdot E_j(\mathbf{r})~ d\mathbf{r}=\delta_{ij}$~\cite{biortho}, the coefficients are computed as follows:
\begin{align}
    c_i^\pm=\int_{\text{OC}_i} n_i^2(\mathbf{r})E_i(\mathbf{r})\cdot E_\pm(\mathbf{r}) ~ d\mathbf{r}\ .
\label{lincomb}
\end{align}
Hence, we can deduce a system of linear equations:
\begin{align}
    \begin{pmatrix}
        k_1-k_\pm & \mu_{12} \\
        \mu_{21} & k_2-k_\pm
    \end{pmatrix}
    \begin{pmatrix}
        c_1^\pm \\
        c_2^\pm
    \end{pmatrix}=0 \ ,
\label{solmu}
\end{align}
    for the unknowns $k_{j}$ and $\mu_{ij}$. Note that, here, we set $k_j$ as free variables and $k_\pm$ fixed, see details in~\cite{RJH}. Eventually, the desired couplings, $\mu_{ij}$, can be obtained as a function of system parameters $R$ and $D$. Figures~\ref{m7m8mu}(a) and (b), respectively, show real and imaginary parts of $\mu_{ij}$, as a function of $R$ for a fixed $D=0.37$. Obviously, it is found that $\mu_{12} \neq \mu_{21}^*$ in this regime of $kr_2\sim 5$.

\begin{figure}[t]
 \includegraphics[width=0.9\columnwidth]{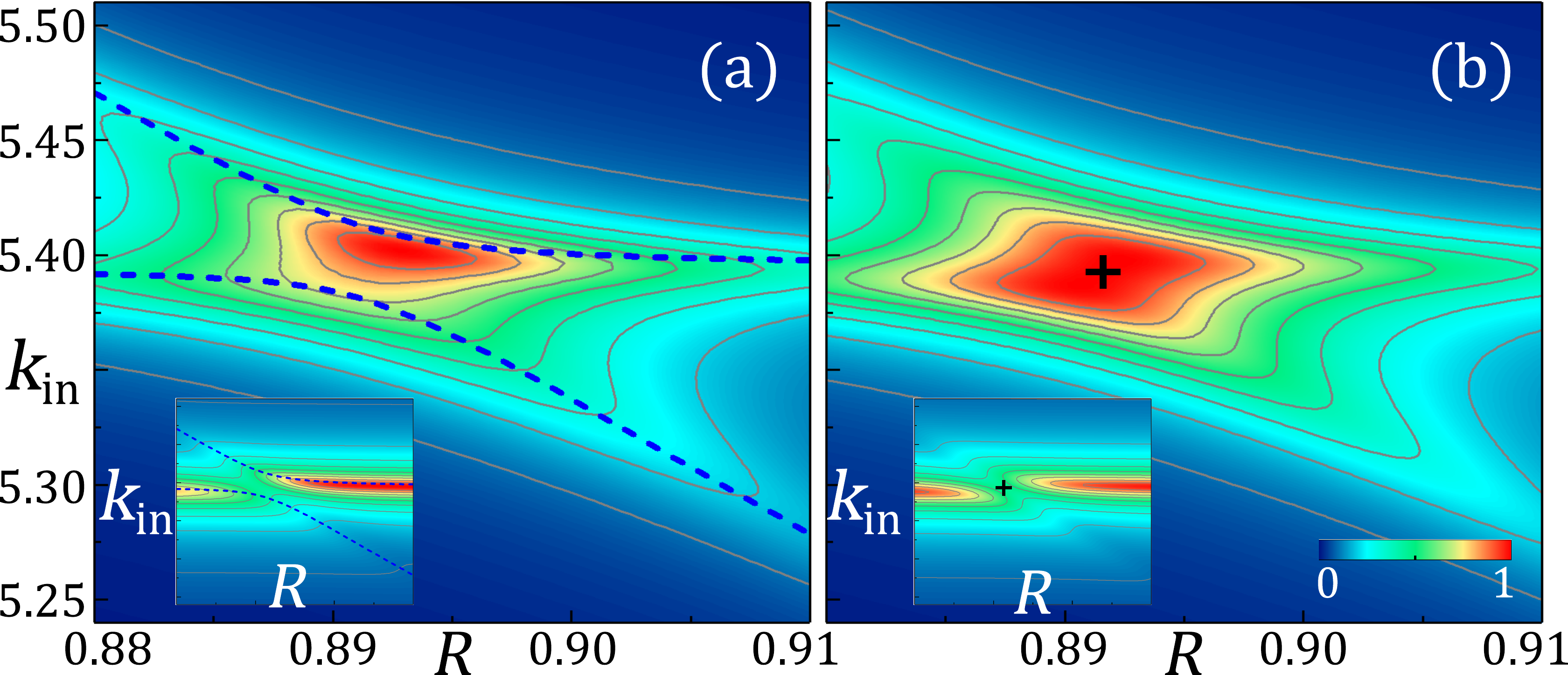}
  \caption{
    (a) and (b) are the TCMT results of EDA spectra of $|a_1|$ with true non-Hermitian and artificial Hermitian couplings, respectively, at $D=0.37$. The insets show those of $|a_2|$. Dashed curves are Re($k_\pm r_2$) obtained by BEM.
  }
  \label{stdampcmt}
\end{figure}

    Inserting the obtained $\mu_{ij}$ into \eq{stdampeq} through the relation $\mu_{ij}\equiv\gamma_{ij}/c$, we calculate R-dependent $|a_1|$ at $D=0.37$ in Fig.~\ref{stdampcmt}(a) ($|a_2|$ in its inset). These figures excellently reproduce the FDTD results given in Fig.~\ref{stddensity}, particularly, the enhanced light transfer around the anti-bonding mode. We can prove that this enhancement originates from the non-Hermitian couplings $\mu_{ij}$ by counter-exemplification of the artificial lossless coupling case; $\mu_{12}=\mu_{21}^*=(\mu_{12}+\mu_{21})/2$. As is shown in Fig.~\ref{stdampcmt}(b) (and its inset), the Hermitian coupling gives rise to a point-symmetric-like feature rather than the enhanced energy density around the anti-bonding mode.

    The effect of the non-Hermitian coupling described above can be understood as a consequence of a decay rate unbalancing around the interaction center.
    Figures~\ref{m7m8mu}(c) and its inset show the imaginary and real parts of $K_\pm$ ($\equiv$ relative eigenvalues for the mean eigenvalue) which are obtained by a direct numerical method (BEM) and by solving Eq.~(\ref{eigen}) with the artificial Hermitian coupling, as well as the true non-Hermitian one.
    In the figures, while the real parts (inset) of them are all almost identical, the imaginary parts show significant differences: the non-Hermitian coupling correctly reproduces the numerical results around the interaction center (shaded area), while the Hermitian one fails it.
    More precisely, the non-Hermitian couplings reproduce the same shift of the branch-cut of the imaginary eigenvalues $(\Delta R)$ from the interaction center (solid orange line). The Hermitian coupling \quoting{never} induce this shift.
    Due to this shift, the bonding and anti-bonding modes at the interaction center happen to have distinctive decay rates [see two big solid dots in Fig.~\ref{eigen}(c)]; longer-lived anti-bonding and shorter-lived bonding modes.
    Because this longer-lived anti-bonding mode has a higher steady-state energy density in the cavity, it can provide much efficient light transfer route than the bonding mode.

    Explanation for this decay rate unbalancing is rather straightforward. Suppose initial states with $\text{Re}(k_1)=\text{Re}(k_2)$ and $\mu_{12}=\mu_{21}^*$, then the difference between two eigenvalues of Eq.~(\ref{eigen}) becomes
\begin{align}
    \Delta k\equiv k_+-k_-=\sqrt{4|\mu_{12}|^2-\text{Im}(\delta k_{12})^2} \ ,
\label{delk}
\end{align}
    where $\text{Im}(\delta k_{12})\equiv\text{Im}(k_1-k_2)$. Though $\text{Im}(\delta k_{12})$ is independent on $D$, the coupling $|\mu_{12}|$ grows from zero to finite value as $D$ decreases from infinity. Accordingly, $\Delta k$ changes from a pure imaginary value to a real number, i.e., weak to strong coupling regime, via EP, at which $4|\mu_{12}|^2=\text{Im}(\delta k_{12})^2$. In this Hermitian coupling case in the strong coupling regime, the interaction center and the point where $\text{Im}(k_+)=\text{Im}(k_-)$, i.e., branch-cut, are identical in $R$ variation for all $D$. That is, the life-times of the bonding and anti-bonding modes at the interaction center are the same, so that the point-symmetric-like feature of EDA in Fig.~\ref{stdampcmt}(b) is induced. In contrast to the Hermitian coupling case, however, if the coupling is lossy, i.e., $\mu_{12}\neq\mu_{21}^*$, $\Delta k$ in \eq{delk} becomes
\begin{align}
    \Delta k=\sqrt{4[(uu^\prime-vv^\prime)+i(uv^\prime+u^\prime v)]-\text{Im}(\delta k_{12})^2}\ ,
\label{diffk}
\end{align}
    for which $\mu_{12}\equiv u+iv$, $\mu_{21}\equiv u^\prime +iv^\prime$, and $\{u^{(\prime)},v^{(\prime)}\}\in\mathbb{R}$. We emphasize that non-zero Im$(\mu_{12}\mu_{21})=uv^\prime+u^\prime v$ shifts the branch-cut from the interaction center, and split the life-times of the doublet modes: longer-lived anti-bonding and shorter-lived bonding mode at the interaction center. Thus, the excitation of the anti-bonding mode is significantly enhanced compared to the bonding mode. The shift direction of a branch-cut can be either to the right or left from the interaction center, so the enhanced mode can be either the bonding or anti-bonding mode, accordingly. In short, the enhancement of light transfer is independent of wavefunction morphologies of modes.
\begin{figure}[t]
	\includegraphics[width=0.9\columnwidth]{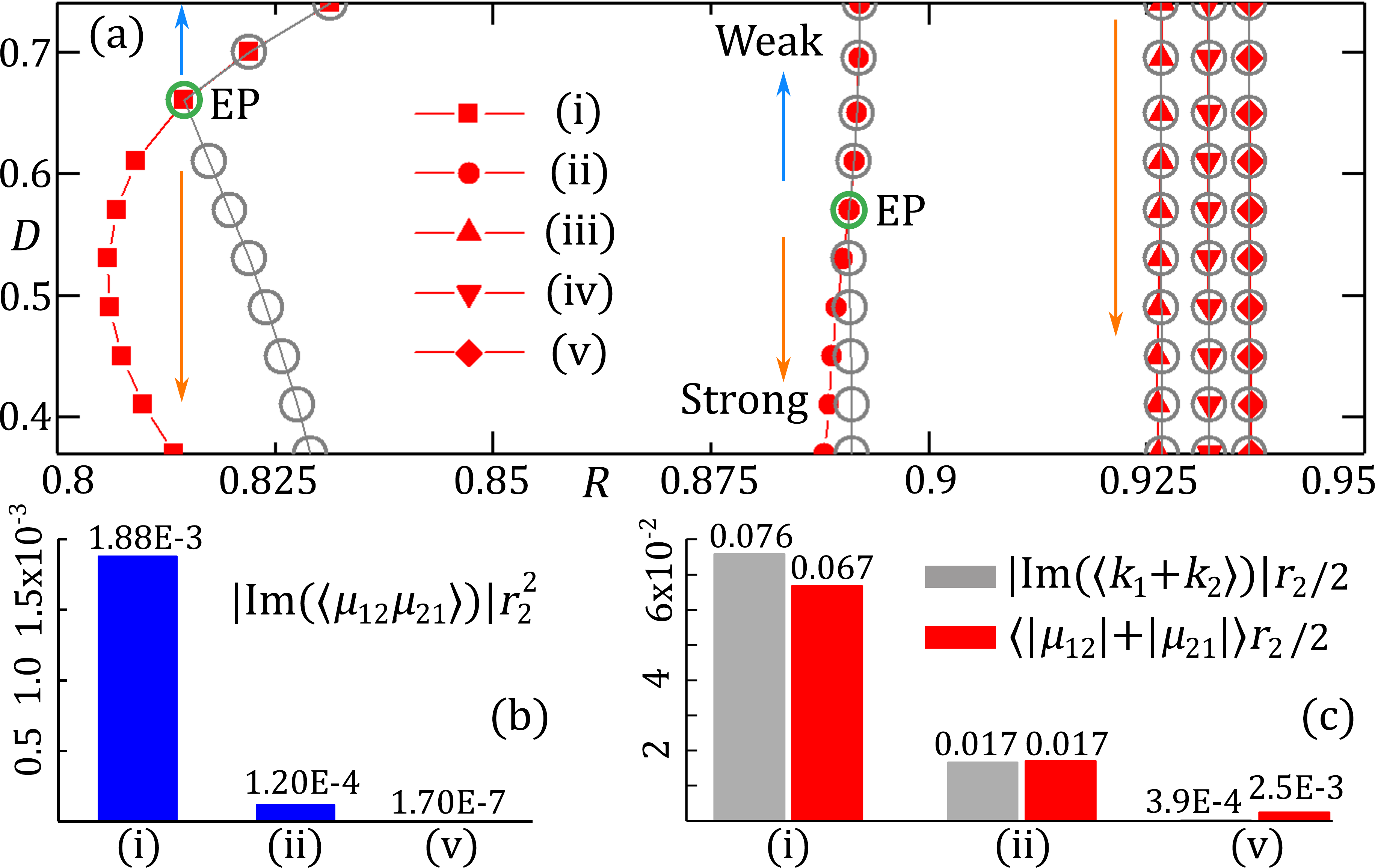}
	\caption{
	(a) The parameter trajectory for the branch-cut (filled symbols) and the interaction center (open circles) for five cases of WGM coupling pairs defined by angular mode numbers: [(i),$(m_1,m_2)=(4,5)$], [(ii),$(7,8)$], [(iii),$(11,12)$], [(iv),$(12,13)$], [(v),$(13,14)$]. (b) shows $|\text{Im}(\langle\mu_{12}\mu_{21}\rangle)|r_2^2$ for the pairs (i), (ii), and (v) obtained at $D=0.37$. (c) is a comparison between the mean values of basis modes' Im($kr_2$) and $\mu_{ij}$ for the same pairs at $D=0.37$.
	} 
	\label{branches}
\end{figure}

    The branch-cut shift induced by the non-Hermitian coupling is found to be much prominent in the quantum regime and diminished as we get into the classical regime. In Fig.~\ref{branches}(a), we obtain the parameter trajectories of the branch-cut and the interaction center in $(R,D)$-space for five different coupling pairs (i)-(v). In the figure, the defined mean wavenumbers $\equiv\text{Re}(\langle k_1+k_2\rangle)r_2/2$ gradually increase from $\sim 3$ to $\sim 8$ for the pair from (i) to (v), where $\langle \cdot \rangle$ denotes averaging over $R$. Clearly, the branch-cut shift from the interaction center for smaller Re($kr_2$) values, e.g., $\sim 3$ for (i), is much substantial, while it is negligible for larger one, e.g., $\sim 7$ or 8 for (iv) and (v).

   In fact, the transition from the non-Hermitian to the Hermitian coupling regime arises very abruptly; it happens far earlier before the classical regime, such as $\text{Re}(kr_2)\sim 8$ in our examples.
   This abrupt transition is readily understood by rapid convergence of Im($\mu_{12}\mu_{21}$) to zero as Re($kr_2$) grows [see, \eq{diffk}].
   In Fig.~\ref{branches}(b), we can see that $|\text{Im}(\langle\mu_{12}\mu_{21}\rangle)|$ is almost zero for (v), whereas it is several orders of magnitude larger than (v) for (i). As the non-Hermitian coupling includes the \quoting{external coupling} via environments, this drastic effect of non-Hermitian coupling profoundly associates with the openness [$\propto$Im$(kr_2)$] of the involved modes.
   Since Im$(\mu_{12}\mu_{21})=uv^\prime+u^\prime v$ is a cross product of real and imaginary parts of $\mu_{ij}$ [see, \eq{diffk}], the order of magnitude of $\text{Im}(\mu_{12}\mu_{21})$ is determined by $|\mu_{ij}|$.
   Now, as $|\mu_{ij}|$ is proportional to $|\text{Im}(kr_2)|$ in Fig.~\ref{branches}(c), we can deduce that the larger decay of basis mode makes stronger inter-site coupling. Therefore, we can conclude here that the openness of basis modes directly promotes both the coupling strength itself and the non-Hermitian coupling effect. That is, the less the mode is confined in the cavity, the higher the impact of the non-Hermitian coupling.
\begin{figure}[b]
	\includegraphics[width=0.95\columnwidth]{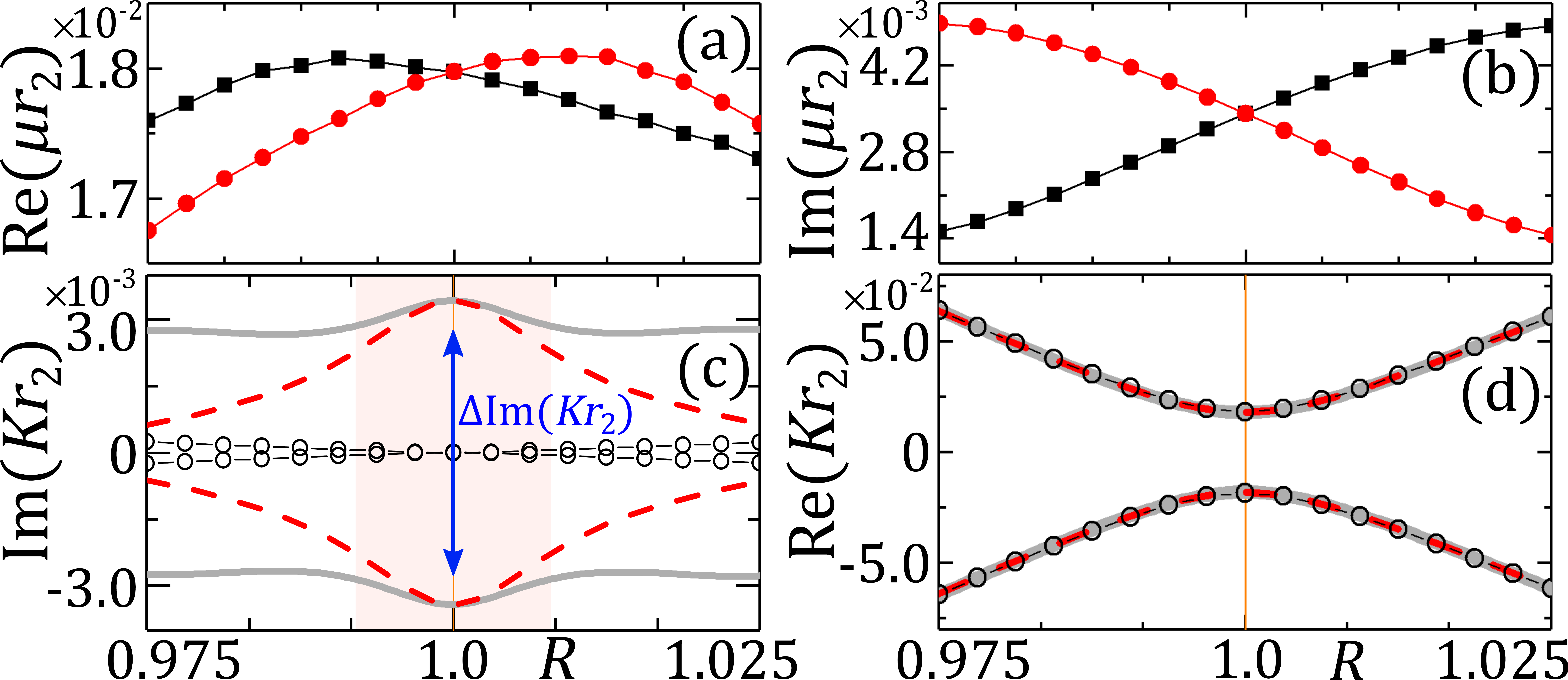}
	\caption{
   Couplings for the pair $(m_1,m_2)=(7,7)$ around $R\approx 1$. Real (a) and imaginary (b) part of $\mu_{12}r_2$ (red-circle) and $\mu_{21}r_2$ (black-square) obtained by \eq{solmu}, as a function of $R$ at $D=0.37$. In (c) and (d), Im($K_\pm r_2$) and Re($K_\pm r_2$) obtained by BEM (gray-solid) are compared to those by \eq{eigen} with the Hermitian (open-circle) and non-Hermitian (red-dashed) couplings. Here, $K_\pm r_2$ are the re-expressed relative eigenvalues for mean eigenvalues. The vertical solid lines in (c) and (d) mark the interaction center.
	}
	\label{R1mu}
\end{figure}

    So far, we have observed that the non-Hermitian coupling has a crucial effect in general inter-cavity mode couplings when the basis modes are confined in non-identical disks with different angular momenta $m$. Here, we remark one special case~\cite{JWRyu}: inter-cavity couplings between the same $m$ modes, $m_1=m_2$, and $R\approx 1$. In Fig.~\ref{R1mu}, we examine the case of $m=7$ and $D=0.37$. Due to the symmetric property for $R$ variation about $R=1$, this COC has balanced values of $\mu_{ij}$ [see, Figs.~\ref{R1mu}(a) and (b)] and a definite interaction center, i.e., it is not dependent on $D$ at all [see, Figs.~\ref{R1mu}(c) and (d)]. Moreover, there is no branch-cut in Riemann-surface; degeneracies in both Re($Kr_2$) and Im($Kr_2$) are always lifted at $R\approx 1$ provided that $D$ is finite~\cite{JWRyu}. In Figs.~\ref{R1mu}(c) and (d), we can find that while the non-Hermitian coupling reproduces this feature accurately, the Hermitian one is valid in Re($Kr_2$) only. The reason for this is exactly the same as before: as Im$(\mu_{12}\mu_{21})=0$ by definition of the Hermitian coupling, the split in Im($Kr_2$) never can be realized, while the non-Hermitian coupling is able to do so. Therefore, again, the broken point-symmetry in EDA spectrum will be induced (not shown) by the non-Hermitian coupling consistently in this system as well.

    To summarize, we have shown that the light transfer in coupled systems can be explained successfully only when the non-Hermitian coupling is considered. We also have found that the non-Hermitian coupling effect dramatically increases as the system size approaches the quantum regime. It has been demonstrated that imaginary parts of a coupling coefficient product, which is responsible for the branch-cut shift, are negligible when the system size is large, yet, they increase rapidly as the system size decreases. This size-dependence effect of non-Hermitian coupling has been proven to associate with the \quoting{opening} of modes. Very recently, several theoretical models have utilized the non-Hermitian coupling to achieve fruitful physical properties, such as hierarchical higher-order exceptional points~\cite{cc1}, unidirectional light propagation~\cite{cc2}, and topological phase transition in resonator array~\cite{cc3}. These ideas are, however, only realizable when we understand genuine features of mode couplings in the nano- and micro-scale technologies, such as silicon photonics and optoelectronic circuits~\cite{Vahala2003,Armani2003}. We believe our results can contribute to developing such devices in the future.
\bibliography{yhhbib}

\end{document}